\documentstyle[11pt]{article}
\def\eqa{\begin{eqnarray}}
\def\eqae{\end{eqnarray}}

\def\tz{\tilde{z}}

\def\del{\partial}
\def\half{\frac{1}{2}}
\def\fracm#1#2{\hbox{\large{${\frac{{#1}}{{#2}}}$}}}
\def\@magscale#1{ scaled \magstep #1}

\catcode`@=11  
\def\un#1{\relax\ifmmode\@@underline#1\else
        $\@@underline{\hbox{#1}}$\relax\fi}
\catcode`@=12

\def\a{\alpha}
\def\b{\beta}
\def\c{\chi}
\def\d{\delta}
\def\e{\epsilon}

\def\g{\gamma}

\def\l{\lambda}
\def\m{\mu}
\def\n{\nu}

\def\p{\pi}
\def\q{\theta}
\def\r{\rho}

\def\t{\tau}

\def\z{\zeta}

\def\S{\Sigma}
\def\U{\Upsilon}

\def\dslash{\not{\hbox{\kern-2pt $\partial$}}}
\def\Dslash{\not{\hbox{\kern-4pt $D$}}}
\def\pslash{\not{\hbox{\kern-2.3pt $p$}}}
 \newtoks\slashfraction
 \slashfraction={.13}
 \def\slash#1{\setbox0\hbox{$ #1 $}
 \setbox0\hbox to \the\slashfraction\wd0{\hss \box0}/\box0 }
 
\font\ro=cmsy10                          
\def\kcr{{\hbox{\ro \char'170}}}                
\def\ktl{{\hbox{\ro \char'170}}}        
\def\ktr{{\hbox{\ro \char'170}}}        
\def\kbl{{\hbox{\ro \char'170}}}        
\def\kbr{{\hbox{\ro \char'170}}}        

\def\plpl{\raise-2pt\hbox{$\raise3pt\hbox{$_+$}\hskip-6.67pt\raise0.0pt
\hbox{$^+$}\hskip 0.01pt$}}
\def\mimi{\raise-2pt\hbox{$\raise3pt\hbox{$_-$}\hskip-6.67pt\raise0.0pt
\hbox{$^-$}\hskip 0.01pt$}} 

\def\bo{{\raise.15ex\hbox{\large$\Box$}}}               
\def\pa{\partial}                                       
\def\TH{{\raise.2ex\hbox{$\displaystyle \bigodot$}\mskip-4.7mu \llap H
\;}}
\def\face{{\raise.2ex\hbox{$\displaystyle \bigodot$}\mskip-2.2mu \llap
{$\ddot
        \smile$}}}                                      

\def\Hat#1{\widehat{#1}}                        
\def\leftrightarrowfill{$\mathsurround=0pt \mathord\leftarrow \mkern-6mu
        \cleaders\hbox{$\mkern-2mu \mathord- \mkern-2mu$}\hfill
        \mkern-6mu \mathord\rightarrow$}
\def\dvec#1{\vbox{\ialign{##\crcr
        \leftrightarrowfill\crcr\noalign{\kern-1pt\nointerlineskip}
        $\hfil\displaystyle{#1}\hfil$\crcr}}}           

\def\fracm#1#2{\hbox{\large{${\frac{{#1}}{{#2}}}$}}}
\def\frac#1#2{{\textstyle{#1\over\vphantom2\smash{\raise.20ex
        \hbox{$\scriptstyle{#2}$}}}}}                   
\def\sfrac#1#2{{\vphantom1\smash{\lower.5ex\hbox{\small$#1$}}\over
        \vphantom1\smash{\raise.4ex\hbox{\small$#2$}}}} 
\def\bfrac#1#2{{\vphantom1\smash{\lower.5ex\hbox{$#1$}}\over
        \vphantom1\smash{\raise.3ex\hbox{$#2$}}}}       
\def\afrac#1#2{{\vphantom1\smash{\lower.5ex\hbox{$#1$}}\over#2}}    

\newskip\humongous \humongous=0pt plus 1000pt minus 1000pt

\newif\ifdtup

\parskip=\medskipamount                 
\thicklines                         

\def\oldheadpic{                                
        \setlength{\unitlength}{.4mm}
        \thinlines
        \par
        \begin{picture}(349,16)
        \put(325,16){\line(1,0){4}}
        \put(330,16){\line(1,0){4}}
        \put(340,16){\line(1,0){4}}
        \put(335,0){\line(1,0){4}}
        \put(340,0){\line(1,0){4}}
        \put(345,0){\line(1,0){4}}
        \put(329,0){\line(0,1){16}}
        \put(330,0){\line(0,1){16}}
        \put(339,0){\line(0,1){16}}
        \put(340,0){\line(0,1){16}}
        \put(344,0){\line(0,1){16}}
        \put(345,0){\line(0,1){16}}
        \put(329,16){\oval(8,32)[bl]}
        \put(330,16){\oval(8,32)[br]}
        \put(339,0){\oval(8,32)[tl]}
        \put(345,0){\oval(8,32)[tr]}
        \end{picture}
        \par
        \thicklines
        \vskip.2in}
\def\oldtitle#1#2#3#4{\oldheadpic\begin{center}\vglue.5in{\large\bf
#1}\\[.6in]
        {#2}\\[.1in] {\it Department of Physics and Astronomy}\\
        {\it University of Maryland, College Park, MD 20742}\\[.6in]
        Physics Publication \#{#3}\\ {#4}\\[1.5in] {\bf
ABSTRACT}\\[.1in]
        \end{center} \begin{quotation}}                 
\def\oldTitle#1#2#3#4#5#6#7{\oldheadpic\begin{center} \vglue .4in
        {\large\bf #1}\\[.4in]
        {#2}\\[.1in] {\it Department of Physics and Astronomy}\\
        {\it University of Maryland, College Park, MD 20742}\\[.1in]
        {#3}\\[.1in] {\it {#4}}\\ {\it {#5}}\\[.4in]
        Physics Publication \#{#6}\\ {#7}\\[.5in] {\bf ABSTRACT}\\[.1in]
        \end{center} \begin{quotation}}                 
\def\border{                                            
        \setlength{\unitlength}{1mm}
        \newcount\xco
        \newcount\yco
        \xco=-21
        \yco=12
        \begin{picture}(140,0)
        \put(\xco,\yco){$\ktl$}
        \advance\yco by-1
        {\loop
        \put(\xco,\yco){$\kcr$}
        \advance\yco by-2
        \ifnum\yco>-240
        \repeat
        \put(\xco,\yco){$\kbl$}}
        \xco=158
        \yco=12
        \put(\xco,\yco){$\ktr$}
        \advance\yco by-1
        {\loop
        \put(\xco,\yco){$\kcr$}
        \advance\yco by-2
        \ifnum\yco>-240
        \repeat
        \put(\xco,\yco){$\kbr$}}
        \put(-20,13){\tiny University of Maryland Elementary Particle
        Physics University of Maryland Elementary Particle Physics
        University of Maryland Elementary Particle Physics}
        \put(-20,-241.5){\tiny  The University of Iowa Particle Theory
        Group The University of Iowa Particle Theory Group The
        University of Iowa Particle Theory Group The University}
        \end{picture}
        \par\vskip-8mm}
\def\bordero{                                           
        \setlength{\unitlength}{1mm}
        \newcount\xco
        \newcount\yco
        \xco=-31
        \yco=12
        \begin{picture}(140,0)
        \put(\xco,\yco){$\ktl$}
        \advance\yco by-1
        {\loop
        \put(\xco,\yco){$\kclr}
        \advance\yco by-2
        \ifnum\yco>-240
        \repeat
        \put(\xco,\yco){$\kbl$}}
        \xco=151
        \yco=12
        \put(\xco,\yco){$\ktr$}
        \advance\yco by-1
        {\loop
        \put(\xco,\yco){$\kcr$}
        \advance\yco by-2
        \ifnum\yco>-240
        \repeat
        \put(\xco,\yco){$\kbr$}}
        \put(-20,12){\ooobacdefghidfghghdhededbihdgdfdfhhdheidhdhebaaahjhhdahbahgdedgehgfdiehhgdigicba}
        \put(-20,-241.5){\oooababaighefdbfghgeahgdfgafagihdidihiidhiagfedhadbfdecdcdfagdcbhaddhbgfchbgfdacfediacbabab}
        \end{picture}
        \par\vskip-8mm}
\def\headpic{                                           
        \indent
        \setlength{\unitlength}{.4mm}
        \thinlines
        \par
        \begin{picture}(29,16)
        \put(165,16){\line(1,0){4}}
        \put(170,16){\line(1,0){4}}
        \put(180,16){\line(1,0){4}}
        \put(175,0){\line(1,0){4}}
        \put(180,0){\line(1,0){4}}
        \put(185,0){\line(1,0){4}}
        \put(169,0){\line(0,1){16}}
        \put(170,0){\line(0,1){16}}
        \put(179,0){\line(0,1){16}}
        \put(180,0){\line(0,1){16}}
        \put(184,0){\line(0,1){16}}
        \put(185,0){\line(0,1){16}}
        \put(169,16){\oval(8,32)[bl]}
        \put(170,16){\oval(8,32)[br]}
        \put(179,0){\oval(8,32)[tl]}
        \put(185,0){\oval(8,32)[tr]}
        \end{picture}
        \par\vskip-6.5mm
        \thicklines}
\def\title#1#2#3#4{\border\headpic {\hbox to\hsize{#4 \hfill UMDEPP#3}}\par
        \begin{center} \vglue .5in {\large\bf #1}\\[.6in]
        {#2}\\[.1in] {\it Department of Physics and Astronomy}\\
        {\it University of Maryland, College Park, MD 20742}\\[1.5in]
        {\bf ABSTRACT}\\[.1in] \end{center} \begin{quotation}}  
\def\Title#1#2#3#4#5#6#7{\border\headpic
        {\hbox to\hsize{#7 \hfill UMDEPP #6}}\par
        \begin{center} \vglue .4in {\large\bf #1}\\[.4in]
        {#2}\\[.1in] {\it Department of Physics and Astronomy}\\
        {\it University of Maryland, College Park, MD 20742}\\[.1in]
        {#3}\\[.1in] {\it {#4}}\\ {\it {#5}}\\[.5in] {\bf ABSTRACT}\\[.1in]
        \end{center} \begin{quotation}}                 
\def\endtitle{\end{quotation}\newpage}                  

\def\ad{{\kern0.5pt \alpha \kern-5.05pt \raise5.8pt\hbox{$\textstyle.$}\kern0.5pt}}

\def\bd{{\kern0.5pt   \beta \kern-5.05pt \raise5.8pt\hbox{$\textstyle.$}\kern0.5pt}}

\def\qd{{\kern0.5pt q \kern-5.05pt \raise5.8pt\hbox{$\textstyle.$}\kern0.5pt}}
\def\Dot#1{{\kern0.5pt {#1} \kern-5.05pt
\raise5.8pt\hbox{$\textstyle.$}\kern0.5pt}}

\begin{document}

\def\gfrac#1#2{\frac {\scriptstyle{#1}}
        {\mbox{\raisebox{-.6ex}{$\scriptstyle{#2}$}}}}
\def\gg{{\hbox{\sc g}}}
\border\headpic {\hbox to\hsize{May 2001 \hfill{UMDEPP 01-000}}}
\par
\par
\setlength{\oddsidemargin}{0.3in}
\setlength{\evensidemargin}{-0.3in}
\begin{center}
\vglue .15in
{\Large\bf Super Gravitons Interacting with the\\ 
Super Virasoro Group{\footnote{Supported  in part by National  
Science Foundation Grant PHY-98-02551.}}  }
\\[.5in]
S. James Gates, Jr.\footnote{gatess@wam.umd.edu}
\\[0.06in]
{\it Department of Physics, 
University of Maryland\\ 
College Park, MD 20742-4111  USA}\\[.1in]
and \\ [.1in] 
V.G.J. Rodgers\footnote{vincent-rodgers@uiowa.edu\,\, }
\\[0.06in]
{\it  Department of Physics and Astronomy, 
University of Iowa\\ 
Iowa City, Iowa~~52242--1479 USA}\\[.9in]
{\bf ABSTRACT}\\[.01in]
\end{center}

\begin{quotation}
{We describe actions that correspond to the interaction of the 
Super Virasoro algebra with supergravitons.  These new field 
theories introduce a superfield that corresponds to dual elements 
of the super Virasoro algebra.  Such elements already appear 
as background fields in the geometric action associated with 
two dimensional Polyakov gravity.   The actions derived here 
supply dynamics to these otherwise background fields.  We are 
also able to extend the definition of these field  theories 
to higher dimensions.  We explicitly exhibit the 2, 3 and 4 
dimensional cases.  Remarkably, the fundamental prepotentials  
describing these dual elements of the super Virasoro algebra,
in each model, agrees with the known prepotentials of the 
corresponding supergravity theory.  These theories might be 
important in the quantization of the super Virasoro group, 
supergravity and in ${\mbox AdS}$ and super ${\mbox AdS}$ 
gravity.}

\endtitle
\noindent

\section{Introduction}

~~~~The Virasoro algebra and the Super Virasoro algebra are the 
underpinnings of string theories and superstring theories.
Representations of these algebras are used in constructing string 
field theories, conformal field theories and low dimensional 
gravitational theories.   Recent work has developed an action 
principle for the dual representation of the Virasoro algebra 
\cite{diff1,diff2} which introduces a rank two tensor, ${\mbox D
}_{A B}$.  One possible interpretation of this field is as a 
``covariant'' background  graviton field.  In two dimensions it 
turns out that, one of the  components of this rank two diffeomorphism 
tensor appears as the background quadratic differential present in 
computations of the two dimensional gravitational anomaly \cite{
polyakov}. Its presence determines the symplectic structure on 
the coadjoint orbit of the Virasoro group \cite{rai,alek} and is 
a signature that the diffeomorphism fields serve as classical 
gravitational fields in two dimensions.  In other words the constant 
quadratic differentials ${b^\star}^\pm$  that are often seen in 
the literature (for example \cite{navarro}),  are precisely the ${
\mbox D}_{\pm \pm}$ components of the diffeomorphism  field in 
light-cone coordinates.  The aforementioned action then, supplies 
dynamics to this field yielding a Virasoro inspired  definition of 
two dimensional classical gravity.  The field equations of the action 
yield constraint equations that are precisely the isotropy equations 
of coadjoint orbits of the Virasoro group where these orbits are  
associated with the two dimensional gravitational anomalies.   
Furthermore the fields ${\mbox D}_{\pm \mp}$ serves as Lagrange 
multipliers for the Gauss' Law generators  that generate  ``+'' and 
``-'' independent coordinate transformations respectively.  

The action constructed in \cite{diff1,diff2} is such that the field
equations for the  ``space-time'' components of ${\rm D}_{A B}$ in
two dimensions become constraints on the initial data.   Explicitly, 
in Minkowski space the field equations for  the ${\rm D}_{0 1}$ 
component dimensions leads to the isotropy equation for the one 
remaining dynamical degree of freedom, ${\rm D}_{1 1}$.  To see 
this we, consider the action
\begin{eqnarray}
\label{action}
S_{\mbox{\tiny diff}}=-&\int& d^nx \sqrt{g}~\fracm1q
\left( X^{L M R}~{\rm D}^A{}_R X_{M L A} +2 X^{L M R} 
{\rm D}_{L A} X^A{}_{R M}\right)\\
-&\int& d^nx \sqrt{g}\left(\frac{1}4  X^{A B}{}_B {} 
\nabla_L \nabla_M{} X^{L M}{}_A+ \frac{\b}2  X^{B G A} 
X_{B G A}\right) ~~~, \nonumber \label{diffeq0}
\end{eqnarray}
where ${\rm X}^{M N R} = \nabla^R D^{M N}$.  Throughout this report we
will reserve the capital Latin letters for space-time indices and
small Greek letters for spinor indices.  In $n$ dimensions, the field 
${\rm D}_{A B}$ has dimensions of $M^{n-4\over 2}$, $\b$ has dimensions 
$M^2$, and $q$ has dimensions $M^{n-8\over 2}$. $\b$ may be interpreted 
as the string tension for a two dimensional theory whereas $q$ is 
determined by the central extension.   Variation with respect to the 
space-time component ${\rm D}_{i 0}$  and setting ${\rm D}_{N 0}=0$ 
leads to the equation,
\begin{equation}
X^{l m 0} \partial_i {\rm D}^{l m} - \partial_m (X^{m l 0} {\rm D}_{l 
i}) -\partial_l(X^{m l 0} {\rm D}_{m i}) - q{~}\partial_i \partial_l 
\partial_m X^{l m 0}= 0 ~~~. \label{diffeq1}
\end{equation}
In $1+1$ this corresponds to the isotropy equation found on the coadjoint
orbit where D corresponds to the quadratic differential.  In other words, 
this field equation becomes $\xi {\rm D}' + 2 \xi' {\rm D} + q{~} \xi''' 
=0$, where the adjoint element $\xi$ corresponds to the conjugate momentum, 
$X^{1 1 0}$, of ${\rm D}\equiv{\rm  D}_{1 1}$.  {\em We interpret this 
as a Gauss Law constraint equation associated with the residual 
time-independent coordinate transformations on the Cauchy data.} Indeed 
in 1+1 dimensions this Gauss' Law constraint is the generator of 
time-independent spatial translations \cite{diff1}.  This constraint 
arises because the space components of  conjugate momentum, $X^{A B}= 
\sqrt{g} \nabla_0 D^{A B},$  transforms under a time independent spatial 
translation as
\begin{equation}
\delta_\xi  X^{a b} = \xi^c \partial_c X^{a b} - X^{c b}\partial_c \xi^a - 
X^{a c}\partial_c \xi^b + (\partial_c \xi^c) X^{c b},
\end{equation}
where $\xi^a$ is the space component a time-independent vector field.
In the 1+1 dimensional case the transformation law for the ``space-space'' 
component of $X^{A B}$ reduces to 
\begin{equation}
\delta_\xi X^{1 1} = \xi (X^{1 1})' - (\xi)' X^{1 1},
\end{equation}
which is the transformation law for adjoint elements of the Virasoro
algebra (modulo central extensions).  Similarly the field $D_{1 1}$ 
transforms as a quadratic differential.  The constraint guarantees 
that the conjugate momentum which lives in the adjoint will not 
transport the Cauchy data into spurious directions. Eq.(\ref{diffeq1}) 
is the higher dimensional extension of this constraint.  For our 
purposes it is only necessary that the one dimensional reduction of 
Eq.(\ref{diffeq1}) reduce to the isotropy equation of Virasoro 
coadjoint orbits.

Although this report is not  interested in geometric action {\em on} 
the coadjoint orbits of the super Virasoro group, it is worth noting 
the appearance of the diffeomorphism field in those cases.  In the 
computation of the two dimensional gravitational anomaly on a cylinder 
with coordinates\footnote{In this chapter, we use $\q$ to describe 
an angular variable.} $(\theta, \t)$, the field ${\mbox D}$ appears 
in the Polyakov effective action as 
\begin{equation}
S = \int d^2 x~{\rm D}(\theta){ \partial_\tau s \over \partial_\theta 
s} + {C \mu  \over 48\pi } \int \,  d^2x ~ \left[ {{\partial^2_{\theta}
s}\over{(\partial_{\theta}s)^2}} \partial _{\tau} \partial_{\theta} 
s -  {{(\partial^2_{\theta}s)^2 (\partial_{\tau} s)}\over{(\partial_{
\theta} s)^3}} \right] ~~~,
\label{EA1}
\end{equation}
and its value governs the symmetries of the symplectic structure for
the field $s(\theta, t)$.   This is to be compared to the geometric 
action for the WZNW model on a cylinder in the presence of a background 
gauge potential  $A_\m$ after gauge fixing,
\begin{eqnarray}
&S& = \int d^2x\, {\rm Tr}\, {\Big \{ } {\rm A}_\theta\, g^{-1}
\partial_\tau  g {\Big \} } + {k \mu} \int \,d^2x ~ {\rm Tr}\,  
{\Big \{ } {\Big (} {{\partial g^{-1}}\over{\partial \tau}}
{\Big )} \, {{\partial  g}\over{ \partial \theta}} \,{\Big \} } 
\nonumber \\
&~&~~~~-~ {k \mu} \int\, d^3x ~ {\rm Tr}\, {\Big \{ } g^{-1} {{
\partial g}\over{\partial \lambda}} \left[  \, {\Big (} {{\partial
g^{-1}}\over{\partial \tau}} {\Big )} \,   {{\partial g}\over{
\partial \theta}} ~-~  {\Big (} {{\partial g^{-1}} \over{\partial 
\theta}} {\Big )} \, {{\partial g}\over{\partial \tau}}
\, \right] \, {\Big \} }  ~~~.
 \label{EA2}
\end{eqnarray}
In this work our efforts will be focused on writing an action that will
complete the picture for the superdiffeomorphism fields that appear 
in the geometric action for the super Virasoro group.

\section{Construction of a Super Diffeomorphism Interaction}

\subsection{Procedural Outline}

~~~~In \cite{diff2} a principle to determine interactions with 
matter fields was developed that was based on the interaction of 
the diffeomorphism field with itself.   There the structure of 
the interaction Lagrangian of the diffeomorphism field was of 
the form
\begin{equation}
{\cal L}_{\rm int} = {\rm X}{}^{L M R}
{\rm Y}{}_{L M R} ~~~,
\end{equation}
where ${\rm X^{L M R}}$ acts as the ``covariantized'' conjugate 
momentum and ${\rm Y_{L M R}}$ is the ``covariantized'' Lie 
derivative of the diff field ${\rm D}_{i j}$, where the small 
Latin indices will denote ``space'' indices and ``0'' will denote 
the time index.  The action in \cite{diff1} and later refine in
\cite{diff2} was based on the observation that the isotropy equations
for the coadjoint orbits can be interpreted as constraints arising
from the time-independent coordinate freedom that still persists after
gauge fixing.  The details of this are explained in references
\cite{diff1,diff2} but we can highlight the salient features of the
construction of the action.  We present an annotated outline of the
construction of the action in what follows.  The construction of 
the covariant interaction Lagrangian went in the following stages.
\begin{enumerate}
\item Contract the conjugate momentum of the field with the 
variation of the field:
\begin{equation}
{\cal L}_{0^{th}}={\rm X^{i j 0}\big(\xi^l \partial_l
D_{i j}+D_{l j}\partial_i \xi^l + D_{i l}\partial_j \xi^l 
\big)} ~~~.
\label{DS1}
\end{equation}
Here we note that this is akin to the pairing of adjoint elements with
coadjoint elements in the geometric actions \cite{rai,alek,loops}.  
This analogy follows since the conjugate momentum $X^{i j 0}$ transforms 
as an adjoint element (see Eqs.(3-4)), while the Lie derivative of 
$D_{i j},$ the term in the parenthesis, transforms as a coadjoint 
element in one dimension.  
\item Replace the fields $\xi^i$ with a space-time component of 
the field, $D^\a_0$:
\begin{equation}
 {\cal L}_{1^{st}}={\rm X^{L M 0}\big(D_0{}^A 
\nabla_A  D_{L M}+D_{A M}\nabla_L 
D_0{}^A + D_{L A}\nabla_M D_0{}^A \big)} ~~~. \label{DS2}
\end{equation}
This is an important ingredient in the construction in it states that
some of the components of the diffeomorphism field, transform in the 
adjoint representation.  Here it is clear that $D^{a}_0$ will
transform as a adjoint element with respect to time independent
spatial translation. Such components serve as Lagrange multipliers
for the constraint equation discussed earlier.  (One possible gauge
fixing condition for the coordinate transformations is to choose
coordinates were $\partial_0 D_{a 0} = 0$.)  Later in the superfield 
actions, we will use a superfield $F^A_0$, that has to leading order 
in its component fields $D^A_0$.   
\item Extend the time directions to covariant directions:
\begin{equation}
 {\cal L}_{\rm int}={\rm X^{L M R}\big(D_R{}^A 
\nabla_A  D_{L M}+D_{A M}\nabla_L D_R{}^A + 
D_{L A}\nabla_M D_R{}^A \big)}~~~. 
\label{KM}
\end{equation}
Here we restore the full general coordinate covariance of the
Lagrangian.  The importance of this will be seen in the supersymmetric 
cases later when we restore the vector indices to the superfield, i.e  
$F^M_0 \,\, \rightarrow  \, F^M_A$
\end{enumerate} 
In \cite{hermann} similar thinking was used to discuss Lagrangians
in which the Cauchy data carries a representation of the Virasoro
algebra.  Perhaps one of the most illuminating comments that we can 
make is to note that this interaction Lagrangian is to be thought 
of as the analog of the first term that appears in either 
Eq.(\ref{EA1}) or Eq.(\ref{EA2}).  In this work we will use this 
principle in order to define a new class of interactions between 
target space supergravity  superfields and the moduli of  Diff 
${\rm S}^1$ gauge fields\footnote{Here we use the word ``moduli'' 
to refer to the parts of the gauge fields with vanishing field 
strengths.}.  These classes of interaction are based on the coadjoint 
representation of the super Virasoro algebra.

\subsection{Superfields and the Super Virasoro Algebra}

~~~~The super Virasoro algebra contains the bosonic Virasoro generators  
$L_m, m\,\epsilon\, {\bf Z}$ and fermionic generators $G_\m, \m\, 
\e\, {\bf Z}$ or ${\bf Z} + \frac{1}{2}$, and a central charge
$\hat{c}$.  It reads
\begin{eqnarray}
\lbrack L_m, L_n \rbrack &=& (m-n) L_{m+n} + \frac{1}{8} \hat{c} 
(m^3-m) \d_{m+n, \,0} \, I ~~~, \nonumber\\
\lbrack L_m, G_\m \rbrack  &=& (\frac{1}{2} m-\m) G_{m+\m} ~~~, 
\nonumber \\
\{ G_\m, G_\n \} &=& -i \,4 \, L_{\m +\n} -i \frac{1}{2} \hat{c} 
\left( \m^2-\frac{1}{4} \right) \d_{\m + \n , \, 0} \, I ~~~.
 \label{SValg}
\end{eqnarray}
A generic element valued in the adjoint of this algebra takes the form
\begin{equation}
{\Hat {\cal A}} ~=~ \sum_{m = - \infty}^{\infty} A^m L_m + 
\sum_{\m = - \infty}^{\infty} A^\m G_\m + \frac{1}{8}a \, \hat{c} 
\, I  ~~~. \label{ele}
\end
{equation}
The parts of the generic elements that do not lie in the center of 
the algebra also may be used to introduce the concept of a superfield
\footnote{The quantity denoted by $\z$ here in Eq.(\ref{SF1}) is simply 
a 1D Grassmann coordinate. In a later section \newline ${~~~~\,}$ we 
use $\q$ to denote the space-time Grassmann supercoordinate variable.
} $A (z,\z)$
\eqa
 && A (z,\z) = \sum_{m = - \infty}^{\infty} (A^m z^{m+1}) + 2 \z 
\sum_{\m = - \infty}^{\infty} (A^\m  z^{\m + \half}) ~~~,
 \label{SF1}
\end{eqnarray}
and the generic element of the algebra written in Eq.(\ref{ele})
has an equivalent representation as a doublet $(A (Z), a)$ with 
$Z= \{z,\z \}$.   We can then derive the following general commutator 
for two adjoint elements 
\begin{equation}
\lbrack [ (A,a) (B,b) ]\rbrack = ( (\partial A) B - A \partial B - i
\half  ({\cal D}A) ({\cal D}B) , \oint dZ (\partial^2 {\cal D}A) \, 
B ) ~~~,
\end{equation}
where $A$ and $B$ are adjoint elements and
\begin{equation}
{\cal D} = \frac{\partial}{\partial \z} + i \z \frac{\partial}
{\partial z} \qquad, \qquad dZ = \frac{dz}{\, 2\p i} d \z ~~~.
\end{equation}
This result is found by first using the representation in Eq.(\ref{ele}),
and calculating the usual bracket using the algebra defined by 
Eq.(\ref{SValg}) \cite{loops,van}. (Note that this definition of
$ {\cal D}$ implies $ {\cal D}^2 = i 2 \pa_z$ which we use below
to derive Eq.(\ref{supercoad}).)

The action on coadjoint elements can be constructed in a similar way 
\cite{rai,van} and we find for the action of a super field $F$ on a 
coadjoint vector $B^\star$ is 
\begin{equation}
\delta_F B^{\star} = -F {\cal D}^2 B^{\star} - \half {\cal D} F 
{\cal D} B^{\star} - \frac{3}{2}{\cal D}^2F B^{\star} + q\,{\cal 
D}^5F ~~~, \label{coad1}
\end{equation}
where $F$ has the decomposition $F=\xi + i \z \epsilon$ and \( B^{\star
}=(u + i \z D,b^{\star}) = (u,D,b^{\star}). \)   Now the isotropy 
equation for the coadjoint element $B^\star$ is given by setting
Eq.(\ref{coad1}) to zero.  In terms of component fields this becomes 
the two coupled equations 
\eqa
-\xi \partial D - \half \epsilon \partial u -\frac{3}{2} \partial 
\epsilon u + b^{\star} \partial^3 \xi -2 \partial \xi D =0 
\label{bosonpart}\\
 -\xi \partial u -\half \epsilon D - \frac{3}{2} \partial \xi u + q\, 
\partial^2 \epsilon = 0 \label{fermionpart}
\eqae
with $ \partial = \partial_z $.

We can make sense out of these transformation rules in higher
dimensions (modulo the central extension).  We replace the 1D Grassmann 
variable $\z$ to a single D-dimensional Majorana spinor $\q^\a$ and the 
supersymmetric covariant derivative operator to 
\begin{equation}
{\cal D}_\m = \partial_\m -\frac i2 \g^N_{\n \m}\q^\n \partial_N  ~~~.
\end{equation}
With this 
\begin{equation}
\{{\cal D}_\m ,{\cal D}_\n \} = -i \g_{\m \n}^N {\partial \over 
\partial z^N} ~~~,
\label{X1}
\end{equation}
where $\g^M_{\mu \nu}$ is determined by the space-time dimension.
With these  $\g^M_{\a \b}$'s  we also introduce $\g^{a \a \b}$ such 
that 
\begin{equation}
\g^{A}_{\a \b} \g^{B \b \l} =\frac12 \delta^\l_\a \eta^{A B} +
\frac12 \S^{A B \l}_\a ~~~,
\label{X2}
\end{equation}
where $\S^{A B \l}_\a$ is anti-symmetric in its space-time indices.  
Then for two dimensions, say,  $F$ is promoted to the vector superfield 
$F^M =( \xi^M + \q^\a \g_{\a \b}^M \epsilon^\b)$ and 
$B^\star$ is promoted to the $\frac32$ spin superfield 
$B_{\m M}=(\U_{\m M} + \q^\a \g_{\a \b}^{N} D_{ M N} + \q^\a 
\q^\b \g^N_{\m [\a } A_{ \b ] M N}) $.  The coadjoint element $D$ 
that appears in $B^{\star}=(u + \z D,b^{\star})$ is the ``space-space'' 
component of $D_{M N}$ when the dimension is two, viz. $D \equiv D_{1 
1}.$ Keeping track of the indices we may write Eq.(\ref{coad1}) as 
\begin{equation}
\delta_{F} B_{\mu M} = F^N \partial_N B_{\mu M} + \partial_M F^N
B_{\mu N} + \frac12 (\partial_N F^N) B_{\mu M} + i~({\cal D}_\l 
F^N \g^{\l \n}_N){\cal D}_\n B_{\m M}. \label{supercoad}
\end{equation}
This is seen as the Lie derivative with respect to $F^M$ on the space-time 
index in the first three summands followed by a supersymmetry transformation 
on $B_{\m M}$ with $\e^\n \equiv ({\cal D}_\l F^N \g^{\l \n}_N)$. This 
combination of a  Lie derivative and supersymmetry transformation is a 
natural extension of the isotropy  equation to dimensions higher than two.   
Note that the superfield $B_{\m M}$ carries a density weight of $\frac12$.
This is consistent with the one dimensional isotropy equation.  Setting
Eq.(\ref{supercoad}) to zero and rewriting in terms of the component 
fields we have 
\eqa
\q^\a \g^A_{\a \m}\big( \xi^M \partial_M D_{A B} + \partial_B 
  \xi^M  D_{C M} +  \partial_M \xi^M D_{A B} + \frac12\eta_{M
  A}~\eta^{C N}~ \partial_N \xi^M D_{C B} - \frac12\partial_A \xi^C 
  D_{C B} \big)&&  \nonumber\\  
  +  \q^\a  \g^A_{\a \r}  \big( \epsilon^\r \partial_A \U_{\m B}+ 
  \partial_B \epsilon^\r \U_{\m M} + \frac12 \partial_A  \epsilon^\r 
  \U_{\m B} -\frac12 ~n~ \epsilon^\r \partial_A \U_{\m B} \big)&& 
  \nonumber  \\ 
  +\xi^A \partial_A \U_{\m B}+ \partial_B \xi^M \U_{\m M} + \frac12
  \partial_M \xi^M \U_{\m A} + i \,n\,\epsilon^\r D_{A B}\g^A_{\m \r} + 
 {\cal O}(\q^2\,\, \& \,\,\S\,\, {\mbox terms})=0,&& \nonumber \\
\label{superisotropy}
\end{eqnarray}
where $n$ in the above is the dimension.  Evaluating this in one 
dimension leads to the Eqs.(\ref{bosonpart},\ref{fermionpart}).

\section{Actions for the Super Virasoro Group}

\subsection{2D and 3D Majorana Spinors}

~~~~We are now in a position to construct an action from the procedure
outlined above.  To begin with we consider the 2D or 3D  Majorana
action since this requires no modification to the superfield $B_{
\mu N}$.   First we define $F^N_A$ as the covariant extension of 
$F^N$ that appears in Eq.(\ref{supercoad}). As stated earlier in the
second part of the procedure of Section(2.1), this should be defined 
in terms of the superfield $B_{\m N}$ since it contains the
diffeomorphism field and its superpartners. The fully covariant
superfield that satisfies the requirements is 
\begin{equation}
F^N_A = E^{\frac12}\, \g^{\a \b}_A {\cal D}_\a B_\b^N.
\label{3Dprep}
\end{equation}
Clearly the leading term of $F^N_0 = D^N_0$ which replaces the vector
field $\xi^i$.  Thus $F^N_0$ serves as the super symmetric extension 
of Eq.(\ref{DS2}) and the covariant extension analogous to Eq.(\ref{KM}).
Next we defined the covariant superfield $B_{A \m M}$ as,
\eqa
B_{A \m M} & \equiv & F^N_A \nabla_N B_{\mu M} + \nabla_M 
F^N_A B_{\mu N} + \frac12 (\nabla_N F^N_A) B_{\mu M} - \frac18
\nabla_{[C} F_{D] A} \Sigma^{ CD \,\l}_{\m} B_{\l M}  \nonumber \\
&{}& ~+ i~({\cal D}_\l F^N_A \g^{\l \n}_N){\cal D}_\n B_{\m M} + q 
~{\cal D}_{\m}\nabla_N \nabla_M F^N_A ~~~,
\end{eqnarray}
where the $\Sigma$ term only contributes to the 3D case. One can 
see that this is the superfield analogue  of the term in parenthesis 
in Eq.(\ref{DS1}) when the subscript $A=0$ above.  Note that $E$ 
is the superdeterminant.  We have also included a term that will 
reproduce the contribution from the central extension when evaluated 
in one spatial dimension.  For the final part we note that $\nabla^A 
B^{\l M}$ will serve as the ``covariantized'' conjugate momentum.  
A suitable kinetic term for the Lagrangian would be
\begin{equation}
\int d^2x \,d \q^\m d \q^\n \, \b\, (\nabla^A 
B_{\mu N})(\nabla_A B_{\n M})\, \eta^{N M}
\end{equation}
where $\b$ is the string tension found in Eq.(1).

With this we can write the action in two dimensions as 
\eqa
S = - &\int& d^2x \,d \q^\m d \q^\n \, \b\, (\nabla^A 
B_{\mu N})(\nabla_A B_{\n M})\, \eta^{N M} \nonumber \\
- &\int& d^2x\, d^2 \q\, \fracm1q \e^{\m \n}\, (\nabla^A B_{
\n B}) B_{A \m M}\, \eta^{B M} ~~~, 
\end{eqnarray}
or
\eqa
S = - &\int& d^2x \,d^2 \q \, \b\, \e^{\m \n} (\nabla^A 
B_{\mu N})(\nabla_A B_{\n M})\, \eta^{N M} \nonumber \\
- &\int& d^2x\, d^2 \q\, \fracm1q \e^{\m \n}\, (\nabla^A 
B_{\n B}) B_{A \m M}\, \eta^{B M}. \label{affirmativeaction}
\end{eqnarray}
Here we have not introduced the determinant $E^{-1}$ in the action 
since our $B_{\n M}$ field is a tensor density of weight $1/2$. 
This action yields the equations of motion and constraint equations
for the 2D super diffeomorphism field.  The constraint equations from
the bosonic field equations of ${\mbox D}_{0 1}$  are the analogs of
the bosonic part of the isotropy equations found on the coadjoint
orbits of the  super Virasoro algebra.  The field equations of the
$\U_\m 0$ and the auxiliary field $A_{\a M N}$ yield the fermionic
constraints on the orbit.

\subsection{2D Chiral Spinors}

~~~~The procedure to build the other supersymmetric actions mimics the
previous example.  One should be careful to choose $F^M_N$ so that is
agrees with Eq.(\ref{DS2}) in the bosonic limit. Furthermore the kinetic 
terms and ``covariantized'' conjugate momenta must respect the Grassmann 
integration. Actions with an odd number of fermions will differ in form 
from Eq.(\ref{affirmativeaction}) since a ``quadratic'' action would 
not be able to recover the proper bosonic limit (Eq.(1)).  The 2D chiral 
case epitomizes these concerns.  

Since the 2D Majorana spinors can be eigenstates of $\g^3$ we can
write spinors in terms of the one component eigenstates of $\g^3$,
${\bar \q^\m}$ and $\q^\m$ corresponding to the ``+'' and 
``-'' eigenvalues.   We write the ``+'' eigenstate superfield
\begin{equation}
{\bar B}_{\m M} = ({\bar \U}_{\m M} + {\bar {\q}}^\nu 
\g_{\m \nu}^N D_{N  M}) ~~~,
\end{equation}
and again define $F_{A B}$ through,
\begin{equation}
 {\bar F}^N_A = \g^{\a \b}_A {\bar {\cal D}}_\a {\bar B}_\b^N.
\end{equation}

Although there is only one spinor component we have purposely
preserved the index structure of the gamma matrices for comparison 
to higher dimensions.  Since we are only integrating over one 
component we must insure that the bosonic sector preserves the 
isotropy equations for orbits of the Virasoro algebra.  The action 
appropriate for chiral fermions in 2D is
\eqa
S = - &\int& d^2x \,d {\bar {\q}}^\m \, \b\, \g^B_{\m \l}\,
\nabla^A {\bar F}_{B M} (\nabla_A {\bar B}^{\l M})\, \nonumber \\
- &\int& d^2x\, d {\bar {\q}}^\m \, \fracm1q \, \g^B_{\m
\l}\,(\nabla^A {\bar F}_{B M} )\,{\bar B}_{A}^{\, \l M}\ + h.c. ~~~. 
\end{eqnarray}
This action, called the affirmative action, gives precisely the one 
dimensional isotropy equations for the super Virasoro sector without 
having to introduce auxiliary fields. It is this action that governs 
the classical dynamics of the quadratic differentials that appear on 
the 2D superstring world sheet.  The vacuum expectation value of this 
field determine the subalgebra  that will be preserved on the world 
sheet and hence govern the vacuum symmetry of the superstring.  We 
are investigating its importance for heterotic string theories.

\subsection{4D Chiral Spinors}

~~~~By using the above 3/2 spin superfield we can make contact with an
interesting chiral variant of this action in four dimensions.  By 
considering a symmetric superfield $B_{\m \a \b}$ field that satisfy 
${\bar {\cal  D}}_\a B_{\m \l \b} = 0$, one has a chiral gravitational 
theory with the action 
\eqa
S = - &\int& d^4x  \,d^2 \q \, \b\, C^{\m \n} (\nabla^A 
B_{\mu \l \r})(\nabla_A B_{\n \a \g})\, C^{\l \a} C^{\r \g}\nonumber \\
- &\int& d^4x\, d^2 \q\, \fracm1q C^{\m \n}\, (\nabla^A 
B_{\n \a \g}) B_{A \m \l \r}\,C^{\l \a} C^{\r \g} + h.c.  ~~~. 
\label{chiral4action}
\end{eqnarray}
The field $d_{\a\b\c\d} = {\cal D_\a} B_{\b\c\d}$ contains the chiral 
graviton ${{\cal D}_{(\a} B_{\b\c\d)}}\mid_{\q=0}$ and a 
Lorentz (1,0) field, ${\cal D^\a} B_{\a \b \d} \mid_{\q=0}$.

\subsection{4D Spinors}

~~~~More interesting is the full non-chiral four dimensional theory.  
Here the relevant superfield is a supervector field akin to the 
field $U_M$ found in supergravity theories related to the 
Einstein-Hilbert action. Let us write our superfield as
\begin{equation}
B_M =  h_M + (\q^\a \U_{\a M} + h.c) + {\q}^\a {\bar \q}^{\dot \a} 
D_{\a {\dot \a} M}  + {\cal{O}}(\q^3),
\end{equation}
where we use ``dot'' and ``undotted'' notation for spinor indices 
when necessary.  The field $F_M^{\,\,N}$ that is suitable for this
case is 
\begin{equation}
 F_{\a {\dot \a}}^{\,\,\,M} \equiv {\bar {\cal D}}_{\dot \a} {\cal 
D}_\a B^M  ~~~.  
\label{4Dprep}
\end{equation}
Note again that the $F^M_0$ component recovers
$D^M_0$ as required in Eq.(\ref{DS2}).
Now since $B_{N}$ is a tensor of rank one and density
zero (i.e. no fermionic indices) we can define
\eqa
B_{{\dot \a} \a N} &\equiv&  F_{\a {\dot \a}}^{\,\,\,M} \nabla_M B_N 
+ B_M \nabla_N F_{\a {\dot \a}}^{~\,M} + \frac{i}2 \, ({\cal D}_\m 
F_{{\dot \a} \a}^{\,\,{\dot \m} \m}){\bar{\cal D}}_{\dot \m}B_N 
\nonumber \\
&{}& +~ \frac{i}2 \, ({\bar {\cal D}}_{\dot \m}F_{{\dot \a} \a}^{\,\,
{\dot \m} \m}){\cal D}_{\m} B_N + q ~\nabla_N \nabla_M F^{\,\,\,M}_{{
\dot \a} \a} ~~~. 
\end{eqnarray}

With this we write
\begin{eqnarray}
S = - &\int& d^4x \,E^{-1}\,d {\bar \q}^2 d \q^2 \, \b\, (\nabla^A 
B_{N})(\nabla_A B_{ M})\, g^{N M} \nonumber \\
- &\int& d^4x\, \,E^{-1}d {\bar \q}^2 d \q^2\, \fracm1q \, (\nabla^A 
B_{B}) B_{A M}\, g^{B M} + h.c. ~~~.
\end{eqnarray}
There is a rather interesting observation to be made about the first 
line above.  Had we begun with the action for 4D, $N$ = 1 supergravity 
used the background-quantum field method and gauge fixed the quantum
supergravity pre-potential, we would arrive at precisely the result on
the first line above. Stated another way, by using solely arguments
about the algebraic structure of the coadjoint super Virasoro representation
we are lead the the gauge-fixed 4D, $N$ = 1 superfield supergravity action 
expressed correctly in term of its prepotential to lowest order in a 
background field formalism.

\section{Conclusion}

~~~~We have constructed an action for the super diffeomorphism field 
that is consistent with the isotropy equations found on the orbits of 
the super Virasoro groups.  In 2D, 3D and chiral 4D, the fundamental 
field is a spin $\frac32$ superfield, $B_{\m N}$ whereas in 4D non-chiral 
we require a spin one prepotential $B_{\a {\dot \a}}$.   This new 
description of the graviton can augment the description of quantum 
gravity in low dimensions since this action supplies dynamics to the
background field that is often called $B$ found in the super geometric 
action \cite{van}
\eqa
S=2q\int d\t\oint dZ\biggl(-\,\half\partial_z^2\tilde{z}D\tz\del_\t
\del_z\tz+\frac{3}{2}\del_z^2\tz D\del_z\tz D\tz \del_\t D\tz && 
\nonumber\\
-\, \del_z^2\tz D\del_z \tz \del_\t\tz+D\del_z^2\tz D\tz
D\del_z\tz\del_\t\tz && \nonumber\\
+\, \del_z^2\tz\del_z\tz \del_\t D\tz-D\del_z^2\tz D\tz \del_\t
D\tz&&\nonumber\\
+\half(\del_z^2\tz)^2D\tz\del_\t\tz + \frac1q B(z) (\del_\t \tilde{
z}/\del_z \tilde{z})\biggr).&&\nonumber\\
\eqae

Furthermore, just as the bosonic diffeomorphism field theory for $D_{A 
B}$ might play an important role in the quantization of the Virasoro 
group and ${\mbox AdS}^3$ gravity so might $B_{\m N}$ in the quantization 
of the super Virasoro Group and  supergravity on ${\mbox AdS}^3$.  In 
fact in \cite{navarro} the coadjoint orbits of the Virasoro group are 
related to the phase space of ${\mbox AdS}^3$ gravity.  However, in 
both the bosonic and supersymmetric cases only certain orbits admit 
K\" ahler structures and hence have a symplectic structure that can 
be geometrically quantized \cite{witten3,rajeev,ramond}.  The origin 
of the obstruction to quantization could be due to the restrictions 
on the diffeomorphism field ${\mbox D}_{A B}$ which foliates the dual 
space into orbits.  Each orbit can be described  by a vacuum expectation 
value of ${\mbox D}_{A B}$ where $< \, {\mbox  D}_{A B} \,> = \delta_A^+ 
\delta_B^+ b^\star,$ where $b^\star$ is a constant.  In general the value 
of $b^\star$ determine the coset space and therefore the global symmetry 
of ${\mbox AdS}^3$ gravity.  One may think of ${\mbox D}_{A B}$ as 
having been ``quenched''. However quantum mechanics may only be 
compatible with certain values of the quenched field.  A more ambitious 
approach to quantization would be to incorporate the interactions of 
${\mbox D}_{A B}$ into the partition function as one does during 
``annealing.''  The annealed theory might admit a consistent quantization 
scheme that allows transitions from the different coadjoint orbits that 
by themselves cannot be quantized. This is exactly analogous to the 
gauged WZNW models\cite{Karabali:1990dk,Bardakci:1988ee,Gawedzki:1989nj,
Karabali:1989au}.  Also the presence of this field will affect the 
number of microstates which could influence the entropy of black 
holes derived from the Virasoro and  super Virasoro coadjoint orbits.

As we have seen for the case of spinors possessing four components, 
the superfield can exist in higher dimensions and extended to several
supersymmetries.  The number of spinors will determine the spin content
of the $B$-superfield prepotential and will interact with metric as a
covariant graviton.   Another interesting extension to the 1+1 dimensional
chiral models would be to use the dual representation of the
geometrically realized super Virasoro algebra that are defined in
\cite{GRn2,gates1}.

Finally, we would like to note a possible additional implication
of this work.  By the proposed method of this work, we see that
there appears a way in which to use the embedding of representations
of the super Virasoro algebra to obtain information directly
about the pre-potential of supergravity theories in greater than
two dimensions.  In our work we have found a curious coincidence
arising from an embedding procedure that leads to identification
between quantities associated with the dual coadjoint elements
on the one side and supergravity pre-potentials on the other
Eq.(\ref{3Dprep}) and Eq.(\ref{4Dprep}).  Should it be possible
to extend this construction universally, it is highly suggestive
that the representation theory of super Virasoro algebras may
be in some way determining the superfield pre-potential structure
of superspace supergravity. These issues are under investigation.

\section{Acknowledgments}

~~~~This work is dedicated to the memory of B.G. Rodgers.  
Also we thank Yannick Meurice and Takeshi Yasuda for discussion. 
This work was supported in part by NSF grant \# PHY-96-43219.
V.G.J.R would like to thank the Mathematical Sciences Research 
Institute in Berkeley, CA for hospitality were some of this
work was completed.

\end{document}